\documentclass[12pt,letterpaper]{article}
\usepackage{lipsum}
\usepackage{graphicx,color}

\usepackage[margin=1.0in]{geometry}

\usepackage[T1]{fontenc}
\usepackage{titling}
\usepackage{authblk}     

\usepackage{titlesec}
\titleformat*{\section}{\large\bfseries}
\titleformat*{\subsection}{\normalsize\bfseries}
\titleformat*{\subsubsection}{\small\bfseries}

\usepackage{indentfirst}
\usepackage{amsmath}
\usepackage{amsfonts,amssymb}

\usepackage{abstract}

\usepackage{fancyhdr}
\title{\bf  \LARGE Effects of Annulation on Low-Reynolds-Number Flows over an Orthocone
}
\author[1]{\bf M. Thakor}
\author[1]{\bf K. H. Seh}
\author[1]{\bf S. R. Gladson}
\author[2]{\bf M. L. Fernandez}
\author[2]{\bf L. C. Ivany}
\author[3]{\bf M. Green}
\author[1]{\bf Y. Sun}

\affil[1]{\small Department of Mechanical and Aerospace Engineering, Syracuse University, Syracuse NY 13244}
\affil[2]{\small Department of Earth and Environmental Sciences, Syracuse University, Syracuse NY 13244}
\affil[3]{\small Department of Aerospace Engineering and Mechanics, University of Minnesota, Minneapolis MN 55455}

\date{}

\begin{document}

\setlength{\headheight}{0pt}
\maketitle

\begin{abstract}
This study numerically examines the influences of transverse annulation around a cone surface on the characteristics of a flow over an orthocone. This work is inspired by \textit{Spyroceras}, a fossilized genus of the nautiloid family during the Paleozoic era, whose method of locomotion is understudied. As a baseline case, a flow over a smooth orthoconic model with a blunt cone end is investigated numerically at Reynolds numbers from 500 to 1500.  As Reynolds increases, two different shedding mechanisms - hairpin-vortex wake and spiral-vortex wake - are captured. We notice that an introduction of annulation over the cone surface changes the critical Reynolds number for the transition of the shedding mechanism. The dominant shedding frequency increases with the Reynolds number for the smooth and annulated cone flows. Moreover, the annulation reduces the dominant frequency for the same Reynolds number and increases the time-averaged drag coefficient. Modal decompositions - Proper Orthogonal Decomposition (POD) and Spectral Proper Orthogonal Decomposition (SPOD) - are used to capture the coherent structures and their oscillating frequencies. We have captured modes corresponding to the hairpin-vortex wake and spiral-vortex wake shedding mechanisms. Comparing the leading POD modes for the smooth and the annulated cone flows, we find that the annulation can reduce the twisting effects of the coherent structures in the wake. Additionally, we find that the SPOD analysis can identify modes presenting both hairpin-vortex wake and spiral-vortex wake in one flow condition as leading modes, while the POD leading modes only reveal one shedding mechanism in each flow. 
\end{abstract}

\section{Introduction}
\label{intro}

\begin{figure} [hbpt]
     \centering
         \includegraphics[width=0.55\textwidth]{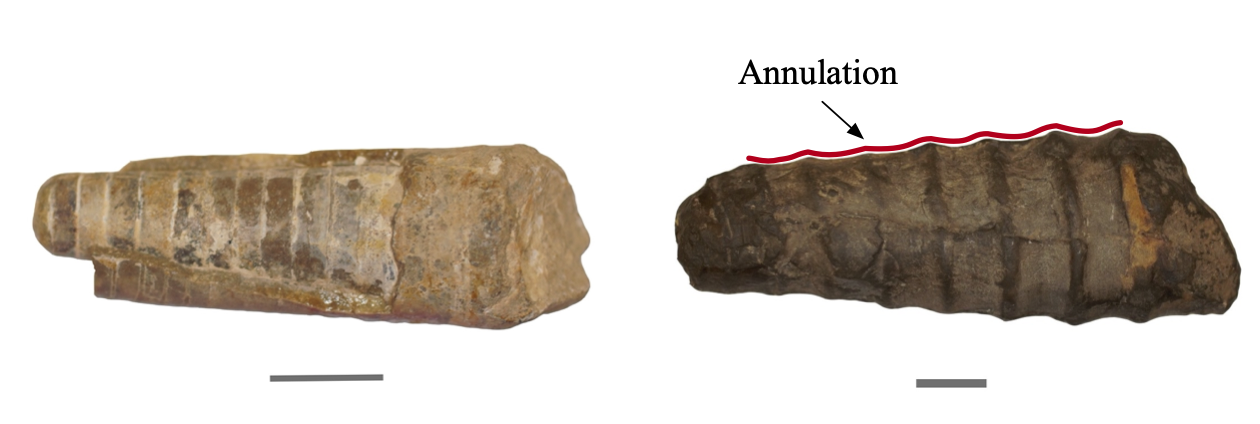}
    \caption{Fossil specimens of various orthocone and surface annulation (scale = 1 [cm]).}
    \label{fig:spyroFossils}
\end{figure}

Adaption through natural selection, while constrained by phylogeny, development, and the mechanical properties of bio-materials, provides an opportunity to discover economical and efficient solutions to meet the functional challenges experienced by organisms daily \cite{baird,briggs}. The form and structure of organisms, including their component parts, offer inspiration and insight for design solutions that may not have direct connections to engineers and scientists. Millions of years ago, \textit{Spyroceras} lived in the ocean, but today, the genus is extinct, and the only close living relative is the \textit{Nautilus} \cite{kroger2008,kroger2009}. A better functional analog might be living squid, but those cephalopods lack a mineralized shell and are more derived in other ways.  Thus, \textit{Spyroceras}'s method of locomotion is still an active area of investigation. Fossilized specimens of the \textit{Spyroceras} shell can be seen in figure \ref{fig:spyroFossils}. Because \textit{Spyroceras} has been extinct for more than 300 million years, the soft body portion and the exact organism locomotion method are unknown \cite{peterman2019}.

The selection of an axisymmetric cone as a baseline to model both the shell's apex (upstream) and end (downstream) is influenced by classical engineering applications and paleontology. When comparing nose cones of trains, it was found that a conical nose resulted in the lowest surface pressure \cite{trainLE} if compared to paraboloids and ellipsoids. Similarly, it has been shown that the trailing edge notably impacts drag production for a bluff body in a flow. Compared to a blunt trailing edge, a sharp trailing edge will produce less drag \cite{naca}. It has also been shown that at a Reynolds number (\textit{Re}) between 10 and 40, rounding off the trailing edge has little influence on the creation of drag \cite{squareCylinderRoundedTE}. Although this result is acquired at a lower \textit{Re}, it poses the potential that for a \textit{Spyroceras} moving at a very low speed, the shape of its soft body may not have an effect. Based on the movement of the modern squid and cuttlefish, it is unlikely that \textit{Spyroceras} would move at such low speeds \cite{peterman}, thus the present work considers a higher \textit{Re} range to simulate more realistic flows over the orthoconic structures.

Flow over a bluff body is an active area of research due to its numerous applications and enriching fundamental fluid dynamics \cite{roshko1993,choi2008,derakhshandeh2019}. In general, the unsteady flow over the cylinders with different cross-sections (circular, square, triangular, rectangular) normal to the streamwise direction is well documented in past studies \cite{tritton1959,schewe2001,agrwal2016,cheng2017}. These studies show that the flow structures and the shedding frequency of the flow primarily depend on the cross-section of the cylinder and the Reynolds number. For flows over an orthoconic structure, extensive studies have been conducted for supersonic and hypersonic flows \cite{anderson1990,gerdroodbary2010}, but the hydrodynamic characteristics of an incompressible cone flow remain unexplored. Hence, this study aims at examining an incompressible flow over an orthoconic structure to fill this knowledge gap.

An educated guess on the organism's movement can be made from modern-day organisms, then verified by utilizing computational fluid dynamics and experimental flow imaging techniques. Many researchers have investigated the impact of shell sculpture in the coiled ammonoid cephalopods, but more work needs to be done to delve into the hydrodynamic characteristics of annulations in straight orthoconic shells. The fossil record shows that \textit{Spyroceras} was a genus of orthocones with transverse annulations on its surface. Studying the hydrodynamic influences of the transverse annulations may provide insights into passive flow control of potential engineering applications in morphological design for micro-aerial and underwater vehicles. 

In what follows, we present the model reconstruction from fossil collection and computational approach with its validation in section \ref{method}. In section \ref{sec:result}, we characterize the three-dimensional flows over the smooth and annulated cones at various $Re$. Moreover, proper orthogonal decomposition and spectral proper orthogonal decomposition are performed on the flow fields to extract the most energetic modes. Summaries are provided in section \ref{summary}.

\section{Methodology and Setup}
\label{method}

We numerically investigate flows over an orthoconic structure resemblant to the general exterior shape of the \textit{Spyroceras} shell. To uncover the influence of the shell's surface annulation on the flow's hydrodynamic features, we will additionally examine flows over a smooth (non-annulated) cone as baseline cases. The flow is characterized by the Reynolds Number, $Re = U_\infty D /\nu\in[500,1500]$, where $U_\infty$ is the free-stream velocity, $D$ represents the diameter of the cone end, and $\nu$ is the free-stream kinematic viscosity.

\subsection{Fossil Reconstruction}
\label{model}

As all the documentations of \textit{Spyroceras} are from the fossil record, a shell reconstruction is performed by measuring the dimensions of a collection of fossil specimens provided by the New York State Museum and augmented by our material. Specimens come mainly from the Middle Devonian Hamilton Group in New York and are about 380 million years old. The main challenges are a flattening of the shells due to burial compression and a lack of complete specimens due to unresolved pre-depositional factors. We assume that the cross-section normal to the streamwise direction was circular originally, while those of the fossils are pseudo-elliptical. Measurements of the major and minor axes of the fossil cross-sections were taken at the peaks and troughs of each annulation (figure \ref{fig:spyroFossils}). These measurements are used to translate the elliptical cross-sections into analogous circular cross-sections. During this process, a cone angle ($\theta_c$) and a length-to-diameter aspect ratio ($L/D$) were approximated. In the present study, the aspect ratio of $L/D=10$ is used.

\subsubsection{Baseline Model} 
\begin{figure}[hbpt]
    \centering
    \includegraphics[width=0.49\textwidth]{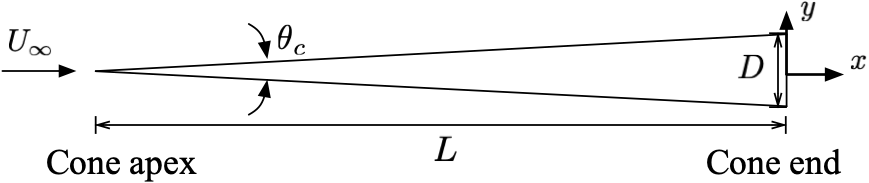}
    \caption{A center $x$-$y$ slice of a smooth cone as a baseline model without surface annulation.}
    \label{fig:baseline}
\end{figure}

As the focus of the present study is to investigate the influences of surface annulation on hydrodynamic features of the flow, a smooth cone model is created for comparison with an annulated model. As shown in figure \ref{fig:baseline}, we use a Cartesian coordinate with the origin at the center of the cone end. The streamwise, normal, and spanwise coordinates are denoted by $x$, $y$, and $z$, respectively. The flow moves from the cone apex on the left towards the cone end on the right.   

\subsubsection{Annulated Model} 
\begin{figure} [hbpt] 
     \centering
         \centering        \includegraphics[width=1\textwidth]{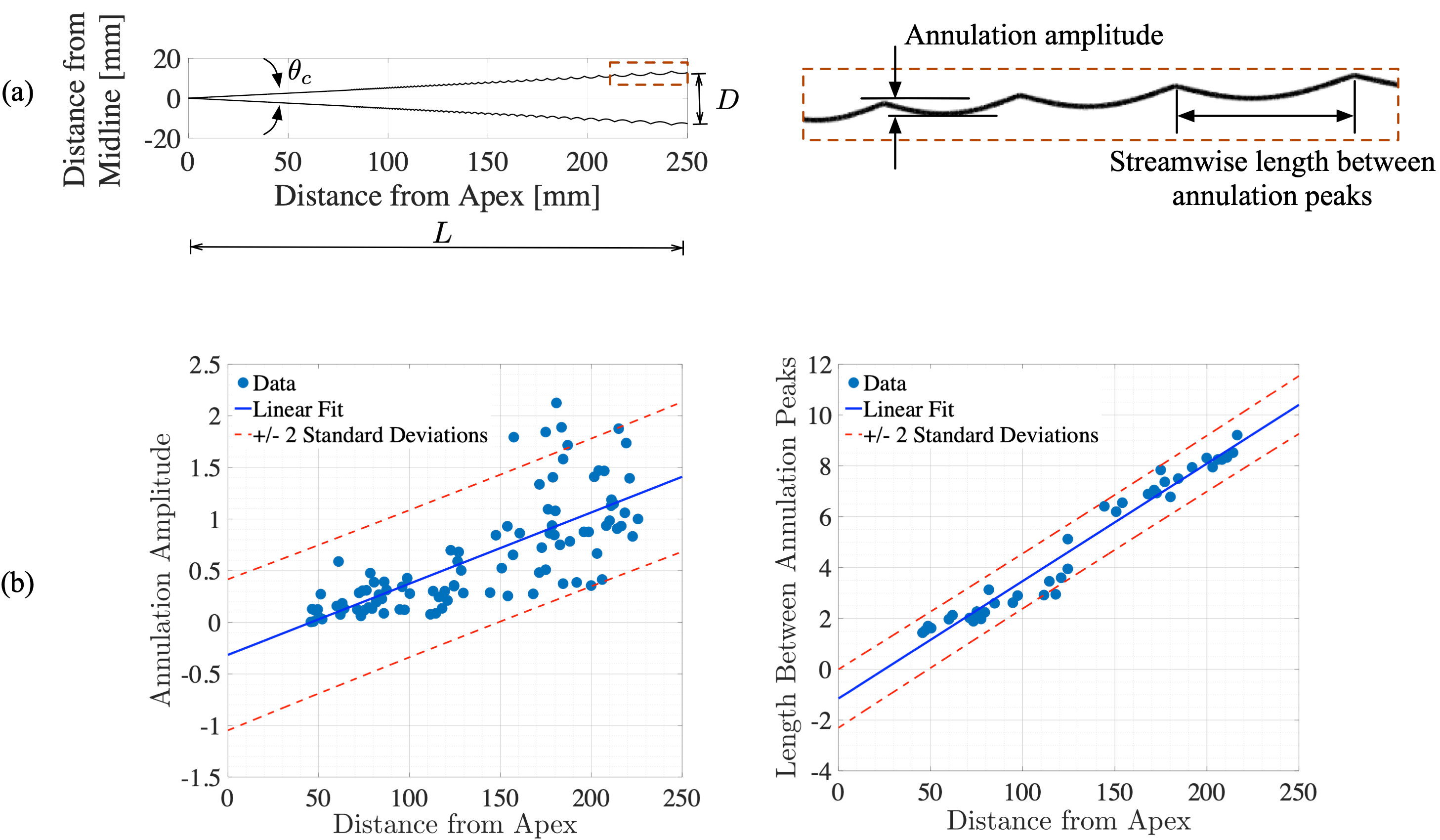}
     \caption{(a) Annulated cone model and (b) linear regressions of measurements of fossil specimens for model reconstruction (all measurements are in a unit of mm). }
     \label{fig:linearReg}
\end{figure}

To quantify the annulation along the cone surface, we use the averaged cone angle to approximate the horizontal distance of each fossil fragment from the apex of the full fossil. As shown in figure \ref{fig:linearReg}(a), we measure the annulation amplitude and the streamwise length between annulation peaks for the model construction. Linear regressions of the annulation amplitude and streamwise length between annulation peaks used for the mathematical model are defined in Eq.(\ref{annulationFunction}) using the sample data set shown in figure \ref{fig:linearReg}, 
\begin{equation}
    y(x) = \begin{cases}
          x/20, ~~~~~~~~~~~~~~~~~~~~~~~~~~~~~~0 < x< L/3 \\
          A(1 - \lvert \sin(x/B) \rvert) + C, ~~~L/3 < x < L 
     \end{cases} 
    \label{annulationFunction}
\end{equation} 
where $A = 0.0101x-0.1367$, $B = 0.000413x+0.0143$, $C = x/20$. The length $L$ is taken as 250 mm for model construction, and the model describes the cone shape in a unit of mm.

\subsection{Numerical Configuration}
\label{sec:num}

Three-dimensional direct numerical simulations (DNS) are performed for incompressible flows over the orthoconic model (discussed in section \ref{model}) using the solver \textit{Cliff} \cite{ham04, ham06} (from the \textit{Charles} software suite \cite{bres, khaligi, khaligi2}), developed by Cascade Technologies, Inc. \textit{Cliff} solves the discretized incompressible Navier--Stokes equations using a node-based second-order finite volume method and second-order temporal scheme. The velocities are denoted by $u$, $v$, and $w$ in $x$-, $y$- and $z$-directions, respectively. We use the cone end diameter ($D$) to non-dimensionalize all lengths, and the length-to-diameter ($L/D$) ratio is maintained at 10 as discussed in the previous section. We note that the non-dimensional coordinate origin $(x/D,y/D,z/D) = (0,0,0)$ is located at the center of the cone end for the numerical study. 

\begin{figure}[hbpt]
    \centering
    \includegraphics[width=1\textwidth]{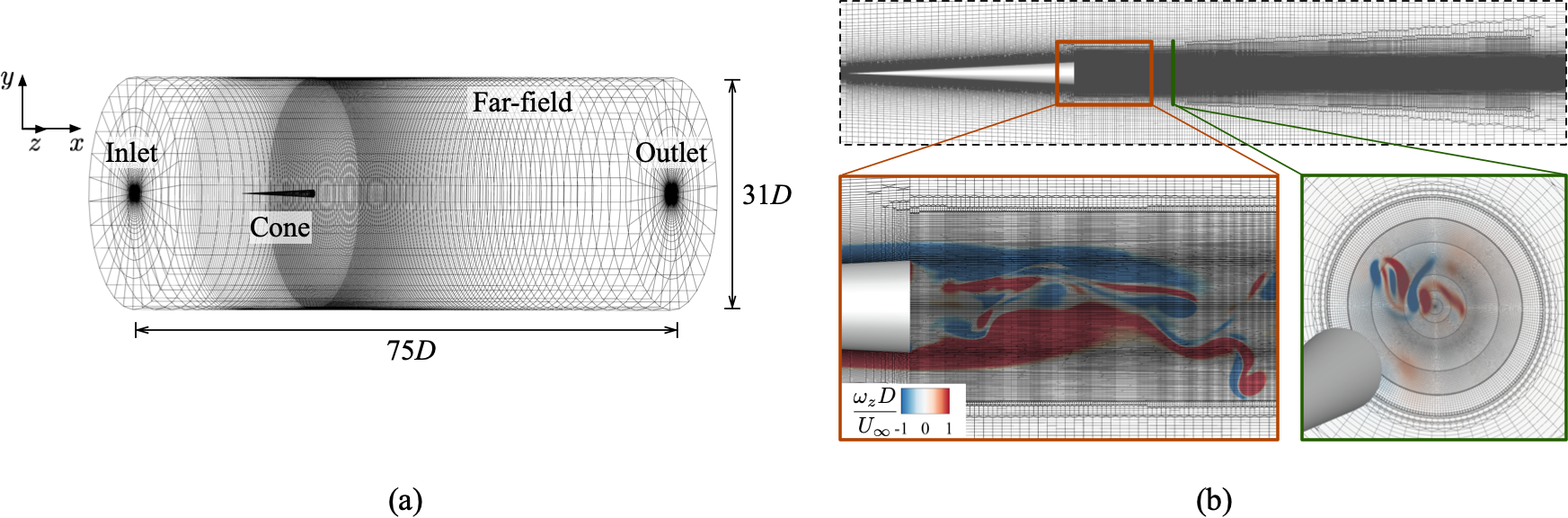}
    \caption{Schematic of computational setup and mesh used for numerical simulations. (a) Computational domain and boundaries; (b) refined mesh around the cone surface and wake area with instantaneous vorticity as background.}
    \label{fig:compDomain}
\end{figure}

The computational domains for both smooth and annulated cases are cylindrical as shown in figure \ref{fig:compDomain}(a). The length and diameter of the domain are $75D$ and $31D$, respectively. At the inlet, a three-dimensional free-stream velocity vector [\textit{u}, \textit{v}, \textit{w}]/$U_\infty$ = [1, 0, 0] is prescribed. The free-stream reference pressure is specified as $P_\text{{ref}} = 0$. A convective outflow boundary condition is specified at the outlet, which allows the wake structures to leave the domain without disturbing the near-field solution. For the far-field boundary, a slip condition is employed. Along the cone surface, a no-slip wall condition is applied. The details of the mesh are shown in figure \ref{fig:compDomain}(b). A structured mesh with non-uniform spacing is prescribed, and refined grid points are concentrated around the cone surface and the wake region to resolve the boundary layer and complex flow structures with small scales. The wall-normal $y^+ \equiv \triangle y u^*/\nu_\infty$ (where $u^* = \sqrt{\tau_0/\rho_\infty}$ is the wall shear velocity, and $\tau_0$ is the wall shear stress) value along the cone surface is less than 1 for all the $Re$ considered. The simulations are performed with time step $\triangle t = 0.002$ for at least 50 convective time, where the convective unit is defined as $L/U_\infty$. Grid information and different versions of meshes are tabulated in table ~\ref{tab:grid_study}.  A grid resolution study has been performed to ensure adequate spatial resolution for smooth and annulated cases. We compared the time-averaged drag coefficient ($\overline{C_D}$) obtained from the coarse and refined meshes as shown in table ~\ref{tab:grid_study}. Although both meshes yield drag coefficients with a negligible difference, the refined mesh can better resolve the small-scale turbulence structures in the wake at higher Reynolds numbers of 1000 and 1500 as shown in figure ~\ref{fig:compDomain}(b). Therefore we use the refined meshes for all the simulations in the present study. 

\begin{table}[hbpt]
\begin{center}
\caption{Comparison of time-averaged drag coefficient $\overline{C_D}$ in grid resolution study. Overbar $\overline{(~)}$ indicates a time-averaged value.}  \label{tab:grid_study}
    \begin{tabular}{cccc|cccc}
        \multicolumn{4}{c}{Smooth Cone} & \multicolumn{4}{c}{Annulated Cone} \\ 
        \hline \hline
        Grid Cells & $Re$ & $\overline{C_D}$ &  $\Delta \overline{C_D}$ & Grid Cells & $Re$ & $\overline{C_D}$ &  $\Delta \overline{C_D}$\\
        \hline
        6.5 million    & 500   & 0.557   & {0\%}      & 4.34 million   & 500  & 0.646   & {0.155\%} \\
        21.9 million    & 500   & 0.557   & -          & 9.2 million   & 500  & 0.647   & - \\
        6.5 million   & 1500  & 0.363   & {0.554\%}  & 4.34 million   & 1500  & 0.397   & {0.253\%} \\
        21.9 million    & 1500  & 0.361   & -          & 9.2 million   & 1500  & 0.396   & - \\
        \hline
    \end{tabular}    
    \end{center}
\end{table}

\subsection{Proper Orthogonal Decomposition}

Proper Orthogonal Decomposition (POD) is a data-driven modal analysis technique to extract the most energetic modes given flow-field snapshots obtained from experimental measurements or numerical simulations \cite{Lumley:ARFM98,taira2017modal}. Snapshot-based POD method is one of the most popular POD analyses. The inputs are snapshots of any scalar (i.e., pressure, temperature) or vector (i.e., velocity, vorticity) field over multi-dimensional discrete spatial points at a discrete time. The snapshots of the flow-field variables are arranged into column vectors after subtracting their time-averaged state, forming a large matrix $Q$ as 
\begin{equation}
    \label{eq:X}
    Q = \begin{bmatrix}         
    \vert & \vert & ~ & \vert  \\ 
    \boldsymbol{q}_1 & \boldsymbol{q}_2  & ... & \boldsymbol{q}_M\\ 
    \vert & \vert &~ & \vert 
          \end{bmatrix},
\end{equation}
where, $Q \in \mathbb{R}^{N \times M}$. The matrix $Q$ consists of $M$ snapshots with each snapshot containing $N$ variables. For fluid flow problems, $N$ is extremely large because it equals the grid points multiplying the number of variables of interest on each point.

If a velocity field is considered for the POD analysis, the resulting eigenvectors $\boldsymbol{\phi}_j$ (also named as POD modes) of the data matrix $QQ^T$ optimally capture the kinetic energy (KE) of the flow, and the resulting eigenvalues $\lambda_j$ represent the amount of the kinetic energy contained by corresponding POD modes. Moreover, these POD modes are orthogonal to each other and are ranked in terms of KE amount. The relation between the original flow field and POD modes is given as follows
\begin{equation}
    \label{eq:qbar}
    \boldsymbol{q}(\xi,t) - \overline{\boldsymbol{q}}(\xi) = \sum_{j} a_j(t) \boldsymbol{\phi}_j(\xi), 
\end{equation}
where, $\boldsymbol{q}(\xi,t)$ is a velocity field with its time-averaged state $\overline{\boldsymbol{q}}(\xi)$, $t$ and $\xi$ denote time and spatial coordinates, respectively. $a_j(t)$ and $\boldsymbol{\phi}_j(\xi)$ represent time-dependent POD coefficients and POD modes, respectively. The temporal coefficients are determined by, $a_j (t)= \langle \boldsymbol{q}(\xi,t),\boldsymbol{\phi}_j(\xi) \rangle$, where $\langle~,~\rangle$ represents an inner product operation. More information about the POD algorithm can be found in the reference \cite{taira2017modal}.

The snapshot-based POD method is widely used as it provides an optimal basis for analyzing unsteady flows; however, such POD analysis lacks information in the frequency domain which is essential for understanding unsteady flow dynamics. To complement the disadvantage of the POD analysis, we will also conduct the Spectral Proper Orthogonal Decomposition (SPOD) analysis in the frequency domain.

\subsection{Spectral Proper Orthogonal Decomposition}

The different forms of spectral analysis \cite{Sieber:JFM16,Rowley:JFM09,towne2018} can identify a basis in which each mode is associated with a single frequency. In the present study, we utilize the SPOD algorithm presented by Towne et al.~\cite{towne2018} to perform spectral analysis. The time-resolved instantaneous states $\boldsymbol{q}(\xi,t)$ of the flow fields are arranged into column vectors after subtracting the time-averaged state $\overline{\boldsymbol{q}}(\xi)$, forming a large data matrix $Q$, similar to the POD data matrix in Eq.(\ref{eq:X}). It is important to note that performing the SPOD analysis requires a much larger number of time-resolved snapshots than the POD analysis.

As described in the study by Towne et al.~\cite{towne2018}, the data set is divided into sequences of blocks. Each block  $Q^{(n)} = [\boldsymbol{q}^{(n)}_1 \boldsymbol{q}^{(n)}_2 {\dots}~ \boldsymbol{q}^{(n)}_{N_s}] \in \mathbb{R}^{N \times N_s}$ contains $N_{s}$ snapshots with an overlap in terms of snapshots with adjacent blocks, here $n$ is the index of the block. A periodic Hanning window is employed over each block to prevent spectral leakage. The Discrete Fourier Transform is performed for each block to obtain Fourier coefficients denoted by $\hat Q^{(n)} = [\hat {\boldsymbol q}^{(n)}_{f_1} \hat {\boldsymbol q}^{(n)}_{f_2} {\dots}~ \hat {\boldsymbol q}^{(n)}_{f_k}]$, where $\hat {\boldsymbol q}^{(n)}_{f_k}$ is the Fourier coefficient at frequency $f_k$ in the $n^{\text {th}}$ block. Next, we rearrange the Fourier coefficients at frequency $f_k$ from each block into a new matrix, $\hat Q_{f_k} = \sqrt{\kappa}[\hat {\boldsymbol q}^{(1)}_{f_k}$ $\hat {\boldsymbol q}^{(2)}_{f_k}$ {\dots} $\hat {\boldsymbol q}^{(N_b)}_{f_k}]$, where $\kappa = \triangle t/s N_b$, $\triangle t$ and $N_b$ are time interval between two snapshots and the total number of blocks, respectively. Here, $s = \sum_{j=1}^{N_s} w_j^{2}$, where $w_j$ are scalar weights used to reduce spectral leakage as the data is non-periodic in each block. The values of $w_j$ are associated with nodal values of a window function. Then we can obtain the cross-spectral density tensor ($S_{f_k}$) at frequency $f_k$ by
\begin{equation}
    \label{eq:cross_spectra}
    S_{f_k} = \hat{Q}_{f_k} \hat{Q}^\ast_{f_k},
\end{equation}  where $^\ast$ denotes the Hermitian transpose. We note that the convergence of the cross-spectral density tensor $S_{f_k}$ depends on the number of blocks ($N_b)$ and the number of snapshots in each block ($N_s$). The values used in the present study are listed in section \ref{sec:POD}. To obtain the SPOD modes for each frequency, we perform an eigenvalue decomposition 
\begin{equation}
    \label{eq:eigen}
    S_{f_k} W \boldsymbol \psi_{f_k} = \boldsymbol \psi_{f_k} \lambda_{f_k},
\end{equation}
where the column vectors $\boldsymbol \psi_{f_k}$ represent SPOD modes and the diagonal matrix of the eigenvalues $\Lambda_{f_k}=\text{diag}(\lambda_1,\lambda_2,...,\lambda_N)$ represents the rank of SPOD modes. The matrix $W$ accounts for the weight and numerical quadrature of the integral on discrete grids. The reader can refer to the reference \cite{towne2018} for the detailed algorithm. 

\section{Results}
\label{sec:result}

In the present study, we characterize flow features at various Reynolds numbers from 500 to 1500. The POD and SPOD analyses are conducted to extract coherent structures highlighting the primary physics of the flows over smooth and annulated orthocones. 

\subsection{Features of Wake Flows}
\label{sec:instant}

\begin{figure} [hbpt]
\centering
    \includegraphics[width=1\textwidth]{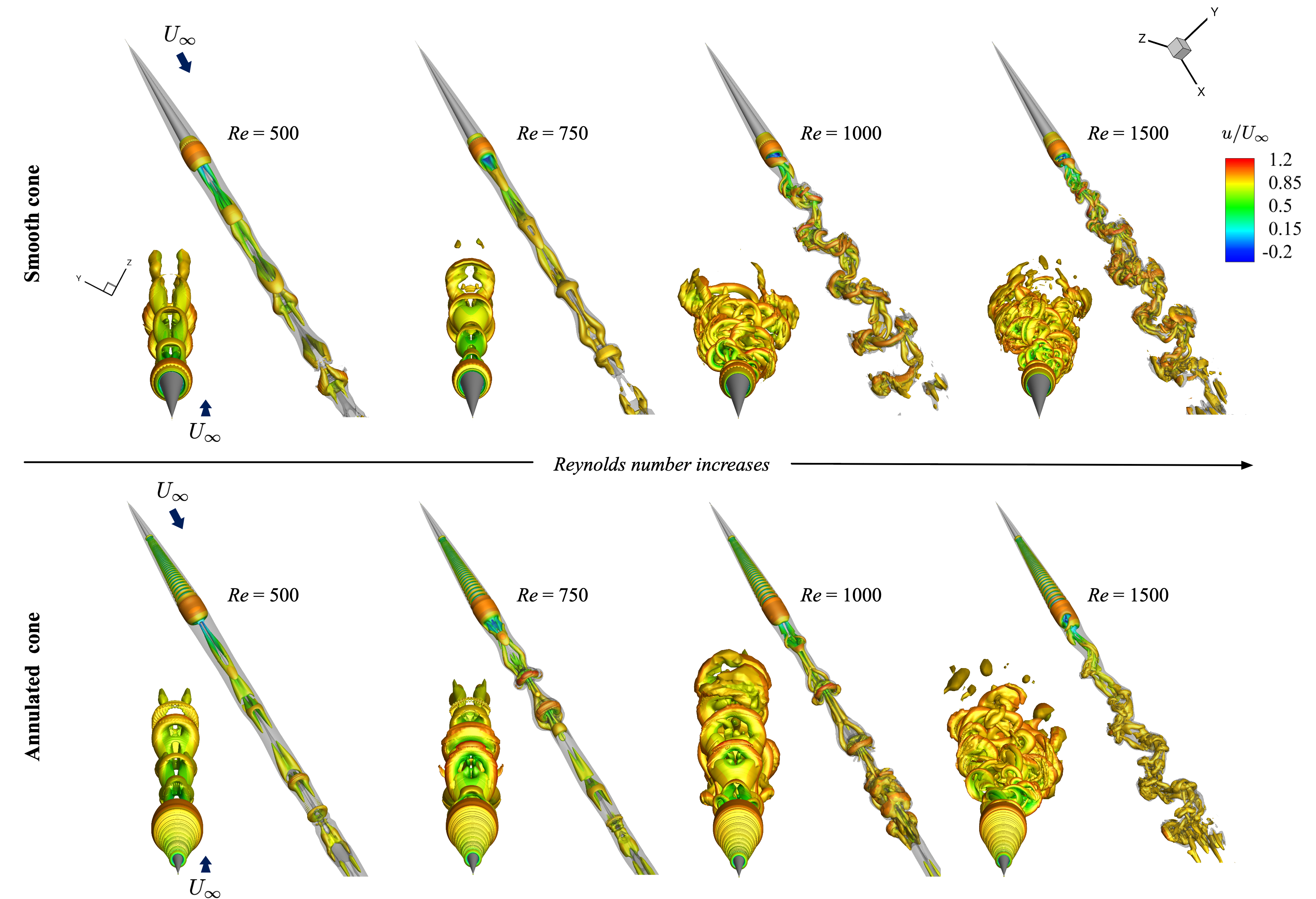}
    \caption{Instantaneous snapshots of flows over the smooth and annulated cones with perspective and front views of iso-surface of $Q$-criterion ($QD^2/U_\infty^2 = 0.01$) colored by streamwise velocity. The transparent grey iso-surface represents vorticity magnitude with $\vert \omega \vert D/U_\infty = 0.25$.}
    \label{fig:isoQ_smooth}
\end{figure}

Representative instantaneous flow fields for smooth cone cases with $Re$ over a range from 500 to 1500 are shown in figure \ref{fig:isoQ_smooth}. Iso-surfaces of $Q$-criterion \cite{Hunt:CTR88} are visualized to highlight the vortical structures in the wake as the flow detaches from the cone end.  At $Re=500$ and $Re=750$, vortical streaks develop and shed periodically from the cone end. When these streaks convect downstream, hairpin vortices grow upon those streaks, which is referred to as {\it hairpin-vortex wake}. However, the flow shedding mechanism and wake flow feature change at higher Reynolds numbers of $Re=1000$ and $1500$, in which the vortical structures formed around the edge of the cone end shed in a rotating manner around the cone axis and convect downstream, creating a spiral flow pattern, which is referred to as {\it spiral-vortex wake}. Moreover, more small-scale turbulent structures are around the spiral vortical structures at higher Reynolds numbers. Hence, the critical Reynolds number for the change of the flow shedding mechanism of the smooth cone flow is between $Re=750$ and $Re=1000$.

Compared to the smooth cone base flows, a significant change in wake flow pattern is observed at $Re=1000$ due to the cone surface annulation (second row in figure \ref{fig:isoQ_smooth}). In the annulated case, the transition from hairpin-vortex wake to spiral-vortex wake happens at a higher Reynolds number between $Re=1000$ and $Re=1500$. Hence the annulation along the cone surface leads to a modification of the flow features in the wake, delaying the appearance of the spiral-vortex shedding phenomenon compared to a smooth cone scenario when the Reynolds number increases.

\subsection{Spectral Analysis of Wake Flows}
\label{sec:frequency}

\begin{figure} [hbpt]
    \centering
    \includegraphics[width=1\textwidth]{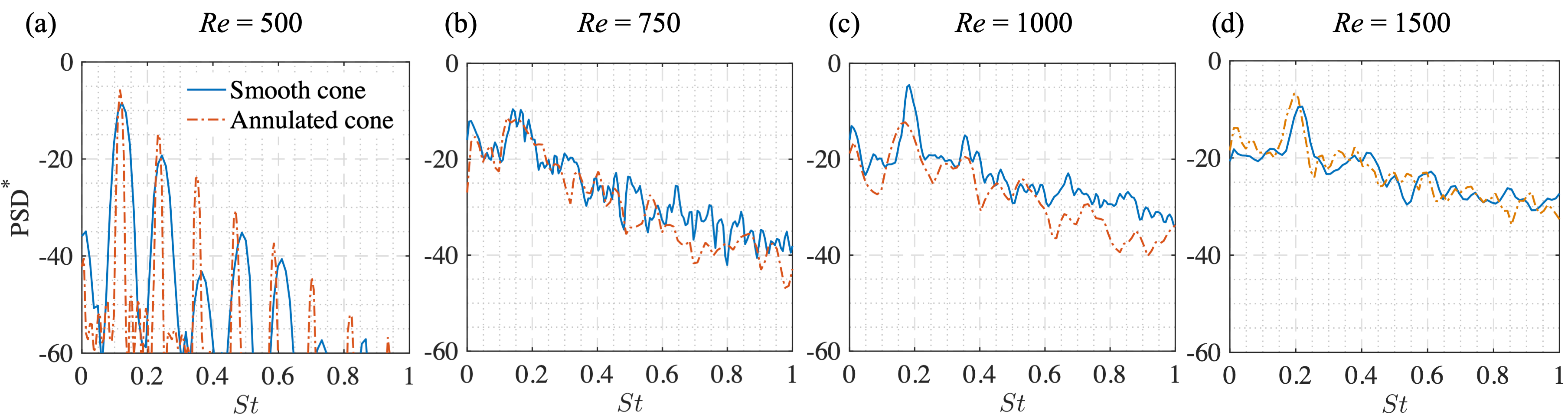}
    ~\\   
   {\footnotesize (e) \begin{tabular}{ccccc}
        $Re$            &500    &750    & 1000  &1500  \\ \hline \hline
       $St$, smooth cone     &0.122  &0.145  &0.179  &0.217  \\
        $St$, annulated cone  &0.118  &0.122  &0.175  &0.200 \\ \hline
    \end{tabular}}
    \caption{Non-dimensionalized Power Spectral Density (PSD$^*$) of streamwise velocity at location of $(x,y,z)/D$ = (5, 0.5, 0) for the smooth and annulated cones flows at (a) $Re=500$, (b) $Re=750$, (c) $Re=1000$, (d) $Re=1500$, and (e) a table of dominant normalized frequencies $St$. }
    \label{fig:PSD}
\end{figure}

Power spectral density (PSD) of the streamwise velocity captured by a probe at the location of $(x, y, z)/D = (5, 0.5, 0)$ is plotted in figure \ref{fig:PSD}. The spectrum is calculated using Welch's method \cite{welch1967} by dividing the time history of the signal into successive blocks with a time step of $\triangle tU_\infty/L = 0.002$ and 75$\%$ overlap Hanning window. The power distribution of the velocity fluctuations in the frequency domain is represented by a non-dimensionalized power spectral density PSD$^*$, 
\begin{equation}
    \label{eq:PSD}
    \text{PSD}^* \equiv 10\log_{10}\frac{U_{xx}}{U_\infty D},
\end{equation}
where $U_{xx}$ is the autospectral density of the probed streamwise velocity. The frequency is normalized as Strouhal number $St = f D/U_\infty$. We also note that we have examined additional four probes at different locations in the wake along the streamwise direction, and similar spectral features are observed. The frequency spectra for $Re = 500$ cases show a clear peak at a dominant frequency, and the following peaks are the harmonic of the dominant frequency, which shows the periodic nature of the flowfield as discussed in section \ref{sec:instant}. For the higher Reynolds number $Re = 1000$ and $1500$, the spectra resemble more broadband due to small-scale fluctuations. Interestingly, the $Re=750$ case also has a broadband frequency spectrum similar to the higher Reynolds number cases, although its wake is closer to the hairpin-vortex wake observed at a lower $Re=500$. In other words, large-scale hairpin-vortex structures can shed from the cone end with intense flow fluctuations. The dominant frequencies (with the highest magnitude) are summarized in a table (figure \ref{fig:PSD}(e)). We find that the flow oscillation frequency increases as the Reynolds number increases, and the increment in the dominant frequency is more significant when the wake flow changes the shedding mechanism from hairpin-vortex wake to spiral-vortex wake.  At the same $Re$, we observe a slight decrement in the dominant frequency by 2-8\% in the annulated cone flows compared to the smooth cone flows. 

\begin{figure} [hbpt]
    \centering
    \includegraphics[width=0.47\textwidth]{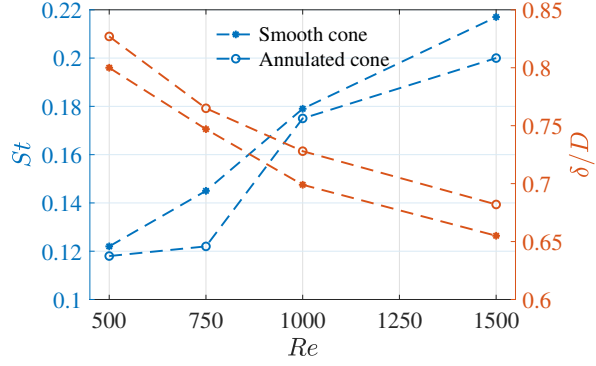}
    \caption{The dominant normalized frequency $St$ (blue) and the time- and azimuthal-direction-averaged boundary layer thickness $\delta / D$ (orange) are shown against the Reynolds number for both smooth and annulated cone flows.}
    \label{fig:BL}
\end{figure}

We further examine the relationship between the dominant frequency in the flows and the corresponding boundary layer thickness at the edge of the cone end in figure \ref{fig:BL}. The time- and azimuthal-direction-averaged boundary layer thickness ($\delta / D$) is calculated at the cone end based on the model described in Eq.(\ref{annulationFunction}). As shown in figure \ref{fig:BL}, the annulation over the cone surface yields an increased boundary layer thickness compared to the smooth cone case at the same $Re$. Accordingly, the flow's dominant frequency decreases. Based on the relation between $Re$ and dominant frequency shown in figure \ref{fig:PSD}(e), the boundary layer thickness at the cone edge in the annulated flow resembles the smooth cone flow associated with a lower $Re$. Hence, the modification of frequencies in the wake flow likely results from changing the boundary layer thickness at the cone end through surface annulation. To further confirm this finding, more datasets and improved accuracy of frequency analysis are necessary.

\subsection{Fluctuations in the Wakes}
\label{sec:fluctuation}
\begin{figure} [hbpt]
\centering
    \includegraphics[width=0.8\textwidth]{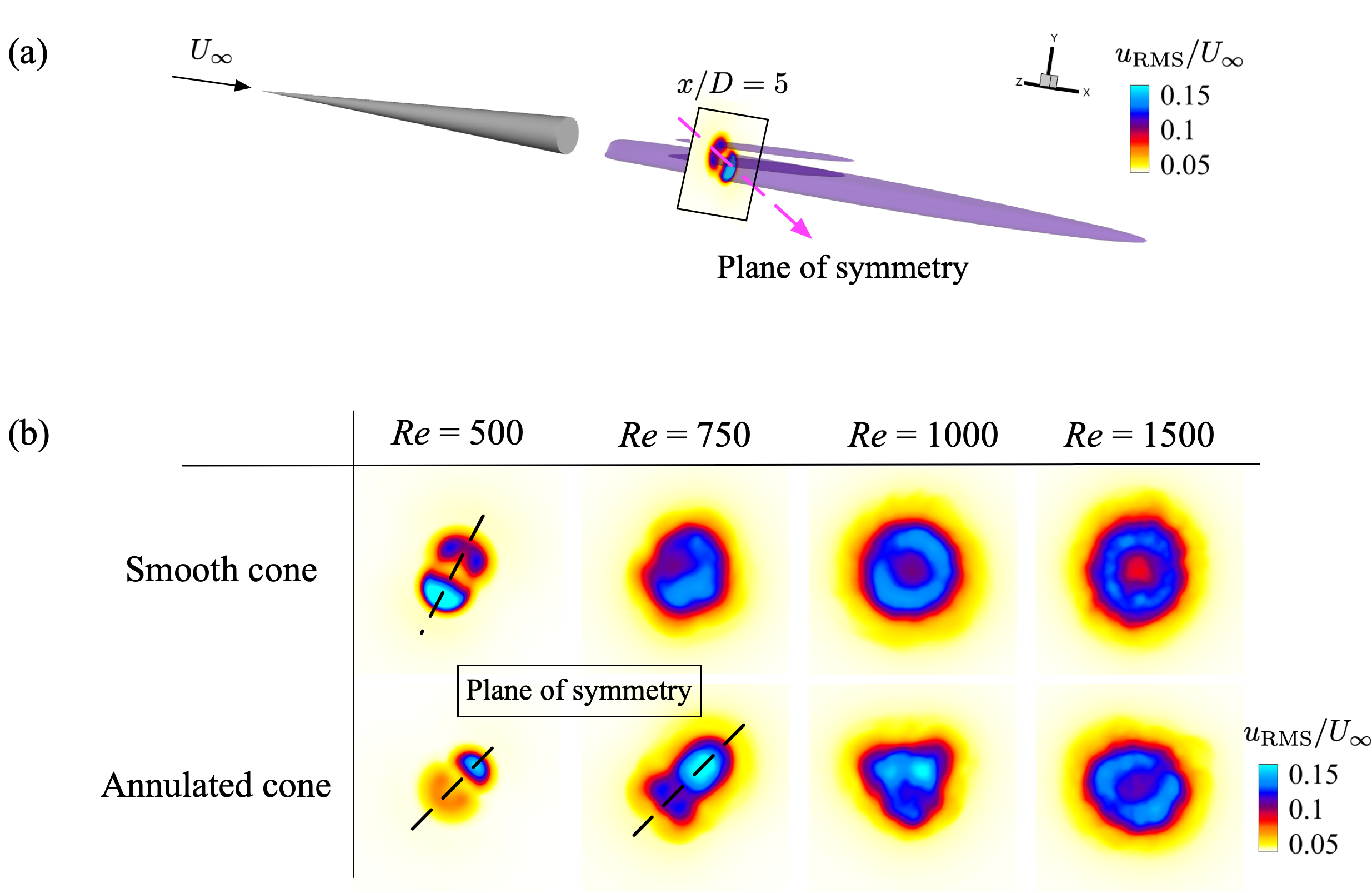}
    \caption{(a) Perspective view of iso-surface of root mean square ($u_\text{RMS}/U_\infty = 0.1$) for smooth cone at $Re = 500$. (b) Root mean square of streamwise velocity ($u_\text{RMS}/U_\infty$) at plane of $x/D = 5$.}
    \label{fig:rms}
\end{figure}

The velocity fluctuations in the wake flows are examined through the root mean square (RMS) of streamwise velocity at the plane of $x/D = 5$ shown in figure \ref{fig:rms}. For the flows presenting hairpin-vortex wake, the hairpin vortex of each flow has a randomly preferred direction to grow in the radial direction while convecting downstream, as we discussed in the instantaneous flow fields. Moreover, the wake flow is symmetric about a plane indicated by a dashed magenta line in figure \ref{fig:rms}(a) and black lines in figure \ref{fig:rms}(b). For the flows presenting spiral-vortex wake, the RMS of streamwise velocity displays an axisymmetric pattern due to the rotating shedding feature observed at higher $Re$ cases. We note here for the annulated cone flow at $Re = 1000$, the RMS of the streamwise velocity profile mixes the features of both wakes because this case exhibits a combined shedding mechanism in which the larger structures present hairpin-vortex shape and the smaller structures present spiral-vortex shape.

\begin{figure} [hbpt]
    \centering
    \includegraphics[width=0.7\textwidth]{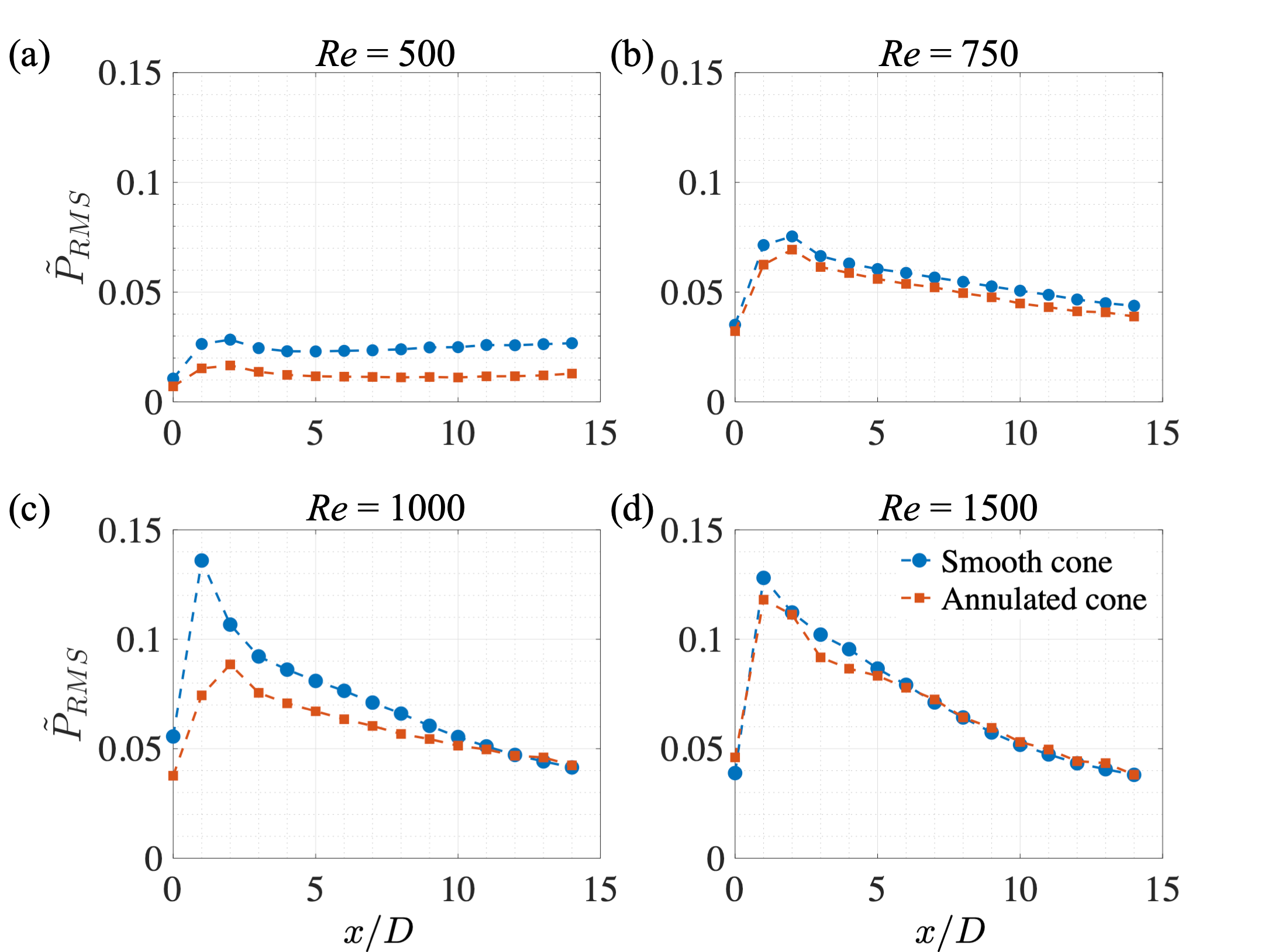}
    \caption{ An integrated $P_\text{RMS}$ is given by $\tilde{P}_\text{RMS} = \int_{S} P_\text{RMS} \,dS $ for an plane of $-5 \le y/D \le 5$ to $-5 \le z/D \le 5$ at multiple streamwise location of $0\le x/D \le 14$. (a) $Re = 500$, (b) $Re = 750$, (c) $Re = 1000$ and (d) $Re = 1500$.}
    \label{fig:pressure_rms}
\end{figure}

Pressure fluctuations on various streamwise-normal planes are investigated for each flow. To quantitatively compare the pressure fluctuation level of the smooth and annulated cone flows, we integrate the RMS of pressure ($P_\text{RMS}$) over an area $S$ ($-5 \le y/D \le 5$ and $-5 \le z/D \le 5$), which is given by $\tilde{P}_\text{RMS} = \int_{S} P_\text{RMS} dS $. As shown in figure \ref{fig:pressure_rms}, the largest pressure fluctuations appear after the cone end between $x/D=1$ and 2 for all the cases, and the fluctuation level decreases as the flows convect downstream, except the case at $Re=500$ that the fluctuation almost remains constant downstream due to low dissipation rate at low $Re$. Comparing the smooth cone and annulated cone cases, the $\tilde{P}_\text{RMS}$ values are lower in the annulated cone flows for all the streamwise locations if both flows have the same type of wake. Moreover, the change in pressure fluctuation due to cone surface annulation is more significant at lower $Re$. At the high $Re=1500$, the difference of $\tilde{P}_\text{RMS}$ is almost negligible between the smooth and annulated cone cases. In general, the annulation over the cone surface weakens the pressure fluctuations in the wake. We also note that as Reynolds number increases, the $\tilde{P}_\text{RMS}$ increases for both the smooth and annulated cone flows.

\subsection{Mean Flow Properties}
\label{sec:mean}

\begin{figure} [hbpt]
    \centering
    \includegraphics[width=0.96\textwidth]{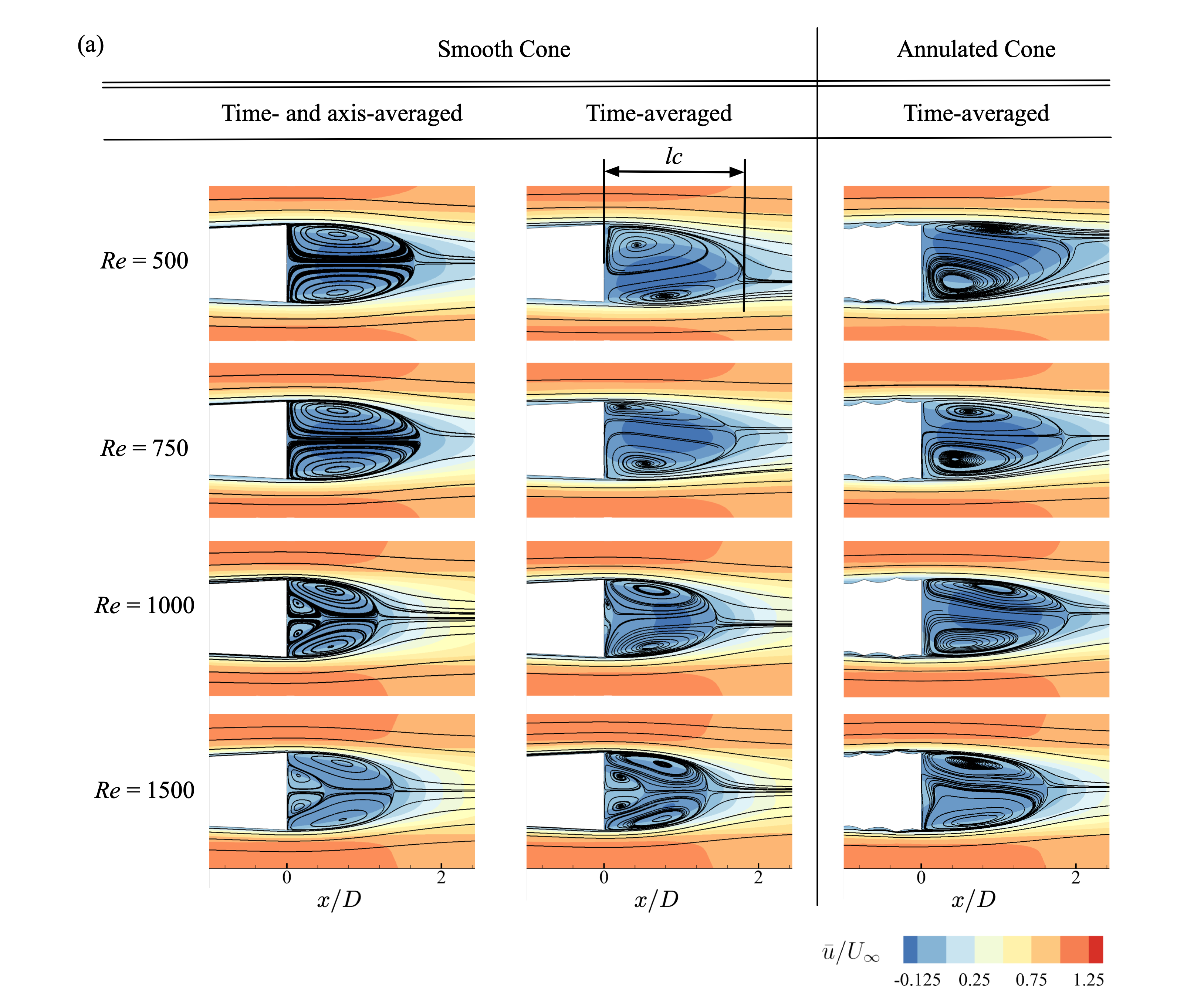}\\
    {\footnotesize (b) \begin{tabular}{ccccc}
        $Re$            &500    &750    & 1000  &1500  \\ \hline \hline
        $l_c/D$ (smooth)     &1.806  &1.717  &1.455  &1.358  \\
        $l_c/D$ (annulated)  &1.927  &1.852  &1.884  &1.642 \\ \hline
    \end{tabular}
    }
    \caption{(a) Mean streamline patterns on the plane of symmetry. The background contours show the corresponding averaged streamwise velocity ($\bar u/U_\infty$), and (b) a table of recirculation length $l_c/D$ for each case. Axis-averaged denotes the azimuthal-direction-averaging procedure.}
    \label{fig:streamline}
\end{figure}

As the hairpin-vortex wake flow is symmetric about a plane shown in figure \ref{fig:rms}, we choose this symmetric plane as a reference location to discuss the time-averaged flows. For the flows with spiral-vortex wake, we analyze both the time- and azimuthal-direction-averaged streamlines and time-averaged only streamlines for the smooth cone cases with corresponding mean velocity ($\bar u/U_\infty$) as background in figure \ref{fig:streamline}(a). As we observe that the hairpin-vortex wake flow is not axisymmetric, the artificial azimuthal-direction-averaging process can yield misleading results. As shown in figure \ref{fig:streamline}(a), if we consider spatial azimuthal-direction-average, in the smooth cone flows with the hairpin-vortex wake at $Re$ = 500 and 750, only one large recirculation zone is formed behind the cone end.
Once the wake flow transits to the spiral-vortex wake at a higher Reynolds number of $Re\ge 1000$, a secondary small recirculation zone appears. However, if we consider only time-averaged flow and choose the plane of symmetry for visualization as shown in figure \ref{fig:pressure_rms}(a), the mean flow pattern is not axisymmetric, and the secondary smaller recirculation zone is not observed in the mean flow pattern at $Re=1000$. Hence, it needs caution to conduct an averaging process of flow over an axisymmetric bluff body because the flow may not present an axisymmetric pattern. This asymmetric mean flow has also been discussed in work by Rigas et al. \cite{rigas16} that the symmetry or axisymmetry flow pattern can break in a three-dimensional flow over an axisymmetric bluff-body. For the flows with the spiral-vortex wake at $Re=1500$, the azimuthal-direction-averaging process yields a similar result to the time-averaged-only flow because of the axisymmetric feature of such a shedding mechanism.

Based on the discussion for smooth cone flows, we only consider the time-averaged only streamlines for the annulated cone flows. The mean flows are asymmetric to the cone axis for all the cases as shown in figure \ref{fig:streamline}. We do not observe small secondary recirculation zones in all the annulated cone cases. The horizontal length of the recirculation zone ($l_c$) is measured as indicated in figure \ref{fig:streamline}, and the values are tabulated in table \ref{fig:streamline}(b). The length of the smooth cone's recirculation zone ($l_c/D$) decreases as the Reynolds number increases. For the annulated cone, the length of the recirculation zone remains similar at $Re\le1000$ and decreases as $Re$ increases to 1500.

The drag of the cone flow is normalized as a drag coefficient $C_D$ defined as  
\begin{equation}
    \label{eq:Cd}
    C_D = \frac{F_D}{\frac{1}{2}\rho_\infty U_\infty^2 D^2},
\end{equation}
where $F_D$ is the integrated drag force over the cone surface. The viscous and pressure drag components are calculated based on the integrated viscous ($F_\text{vis}$) and pressure ($F_\text{pre}$) forces over the cone surface, respectively. The drag force is the summation of the viscous and pressure forces $F_D = F_\text{vis} + F_\text{pre}$.  We investigate the time-averaged drag coefficient ($\overline{C_D}$) and contributions made by viscous force ($\overline{C_{D, \text{vis}}}$) and pressure force ($\overline{C_{D, \text{pre}}}$) components, and their values are tabulated in table \ref{tab:drag}. For both the smooth and annulated cones flows, the viscous force ($\overline{C_{D, \text{vis}}}$) becomes less dominant as the Reynolds number increases. The overall drag increases as the annulation are introduced to the cone surface, but the changes in each drag component (viscous and pressure forces) are different. The viscous force decreases ($\Delta \overline{C_{D,\text{vis}}}<0$) with annulation for all the unsteady cases with $Re\ge500$. Because the flow trapped in the groove of the annulations allows the external flow to pass with a slip-like boundary condition over the cone, it reduces viscous force along the cone surface. In the smooth cone flow, the external flow directly contacts a no-slip surface, which yields a larger viscous force. While examining the relation between pressure drag and wake mechanism, we found that the pressure drag coefficient marginally decreases as $Re$ increases regardless of the cone surface geometry if the flows have the same wake mechanism. However, a sudden increase in the pressure drag coefficient appears when the wake mechanism changes from the hairpin-vortex wake to the spiral-vortex wake, which indicates that the spiral-vortex wake is associated with a higher pressure drag around the wake-transition condition. Due to the limited cases considered in the present work, a further increment in the $Re$ for the spiral-vortex wake flows or a further decrement in the $Re$ for the hairpin-vortex wake flows may break this relation. In other words, the type of wake mechanism is not the only factor determining the range of the pressure force drag component. Furthermore, at the same $Re$ with the same wake mechanism, the pressure drag increases significantly due to the annulation, which suggests that the annulation over the cone surface plays a primary role in increasing the pressure drag component.

\begin{table}[hbpt]
\begin{center}
\caption{Comparison of individual components ($\overline{C_{D,\text{vis}}}$ and $\overline{C_{D,\text{pre}}}$) of time-averaged drag coefficient $\overline{C_D}$.}
{\footnotesize
    \begin{tabular}{c|ccc|ccc|ccc}
        \multicolumn{1}{c}{} & \multicolumn{3}{c}{Smooth Cone} & \multicolumn{3}{c}{Annulated Cone} & \multicolumn{3}{c}{Difference}\\ \hline \hline
        Re & $\overline{C_{D,\text{vis}}}$ & $\overline{C_{D,\text{pre}}}$ &  $\overline{C_D}$ & $\overline{C_{D,\text{vis}}}$ & $\overline{C_{D, \text{pre}}}$ &  $\overline{C_D}$ & $\Delta \overline{C_{D,\text{vis}}}$  &  $\Delta \overline{C_{D,\text{pre}}}$ &  $\Delta \overline{C_{D}}$\\
        \hline
        500     & 0.480   & 0.077   & 0.557     & 0.479   & 0.167  & 0.646   & {-0.20\%}   & {116.8\%}   & {15.97\%}\\
        750     & 0.376   & 0.075   & 0.451     & 0.364   & 0.151  & 0.515   & {-3.19\%}   & {101.3\%}   & {14.19\%}\\
        1000    & 0.324   & 0.119   & 0.443     & 0.303   & 0.150  & 0.453   & {-6.48\%}   & {26.05\%}   & {2.25\%}\\
        1500    & 0.254   & 0.106   & 0.360     & 0.233   & 0.163  & 0.396   & {-8.26\%}   & {53.77\%}   & {10.00\%}\\
        \hline
    \end{tabular}
    }
    \label{tab:drag}
    \end{center}
\end{table}

\begin{figure} [hbpt]
    \centering
    \includegraphics[width=1\textwidth]{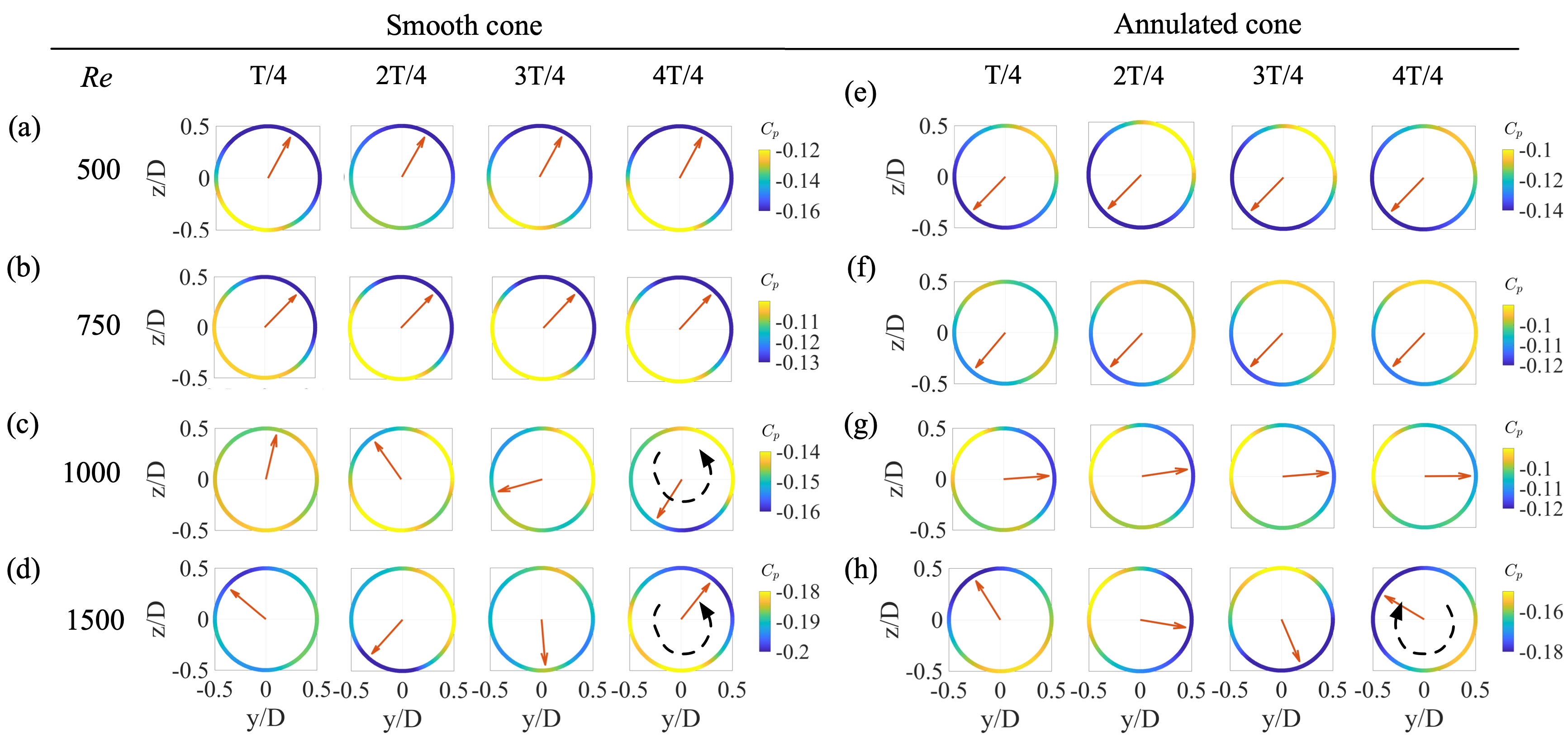}
    \caption{Instantaneous pressure coefficient ($C_p$) at the cone end ($x/D = 0$) with equal time interval ($T/4$). The arrow indicates the integrated pressure force direction. Note that the length of an arrow does not represent the magnitude of the pressure force. (a)(e) $Re = 500$, (b)(f) $Re = 750$, (c)(g) $Re = 1000$, and (d)(h) $Re = 1500$.}
    \label{fig:smooth_pressure}
\end{figure}

The behavior of the surface pressure near the cone end ($x/D=0$) is characterized by examining the instantaneous pressure coefficient, $C_p = (P-P_\text{ref})/0.5\rho_\infty U_\infty ^2$, as shown in figure \ref{fig:smooth_pressure}. The time interval $T/4$ is derived based on the dominant frequency ($1/T$) of each case calculated from the PSD analysis (figure \ref{fig:PSD}). The red arrow indicates the direction of the integrated instantaneous pressure force around the cone end surface. We note that the length of the arrow does not represent the magnitude of the pressure force. As shown in figure \ref{fig:smooth_pressure}(a)(e) and (b)(f), the direction of the pressure force is invariable for a one-period interval ($T$) for both smooth and annulated cone flows at low Reynolds numbers ($Re \le 750$). As discussed earlier, at lower Reynolds numbers ($Re \le 750$) flows, both the smooth and annulated cone flows exhibit the hairpin-vortex wake feature in which the vortices grow in a particular direction; hence the direction of the surface pressure force in these cases remains unchanged.

For the high Reynolds number ($Re \ge 1000$) flows in which spiral-vortex wake is dominant, the direction of the integrated pressure force moves counterclockwise in the smooth cone flows (figure \ref{fig:smooth_pressure}(c) and (d)) and clockwise in the annulated cone flow (figure \ref{fig:smooth_pressure}(h)) in time as indicated by a black dashed arrow. We note at $Re=1000$, there is a significant change in the behavior of the surface pressure at the cone end between the smooth and annulated cone flows. The direction of the pressure force rotates counterclockwise in the smooth cone flow (figure \ref{fig:smooth_pressure}(c)), whereas it is unvaried for the annulated cone (figure \ref{fig:smooth_pressure}(g)). This change of integrated pressure force direction corresponds to the shift of the shedding mechanism at $Re=1000$ due to the annulation effect. As shown earlier in the instantaneous flow fields that the flow sheds in a combined mechanism at $Re=1000$ with annulation, the wake near the cone end exhibits a feature closer to the hairpin-vortex wake, hence the integrated pressure direction remains unvaried.

\subsection{Modal Decomposition}
\label{sec:POD}

\begin{figure} [hbpt]
    \centering
    \includegraphics[width=0.85\textwidth]{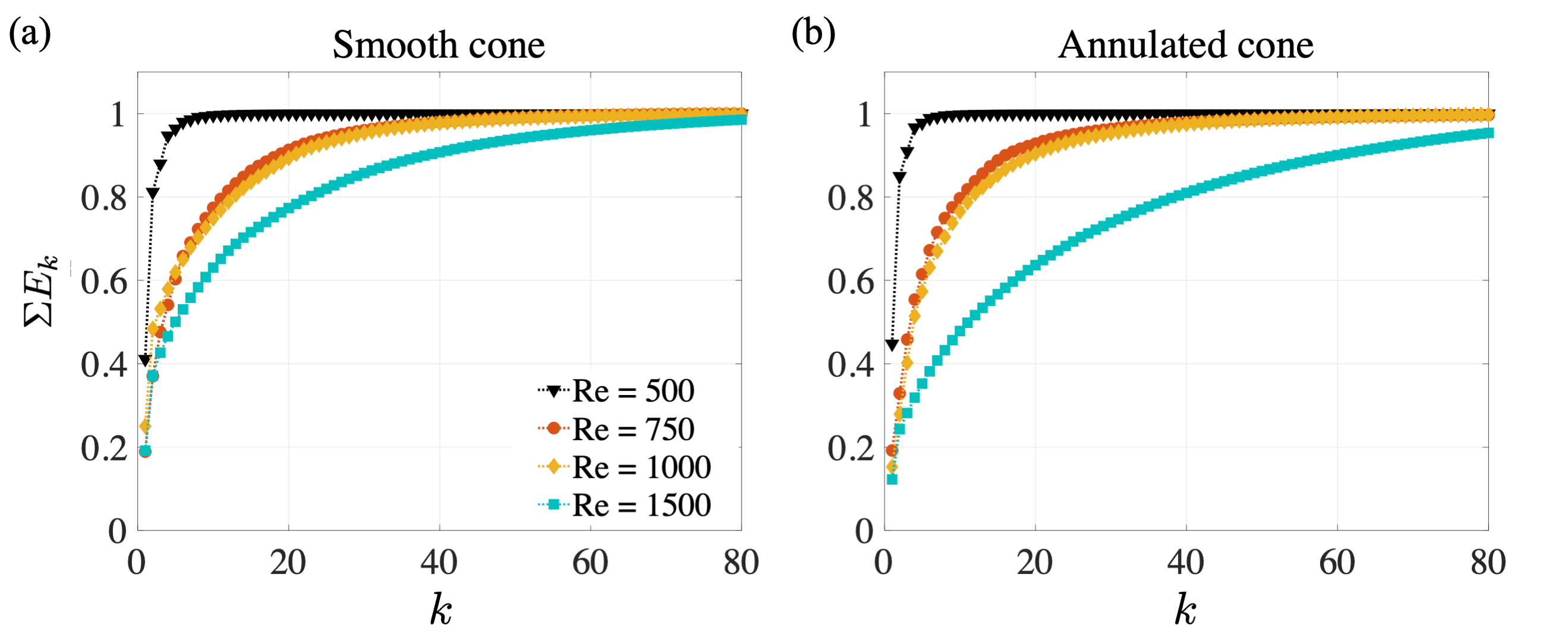}
    \caption{Ratio of the sum of the energy of the POD modes to the total energy of all the POD modes ($\sum E_k = \sum_{j=1}^{k} \lambda_j/\sum_{j=1}^{n} \lambda_j $). (a) Smooth cone and (b) annulated cone flows.}
    \label{fig:energymode}
\end{figure}

We perform POD analysis using the velocity flow fields to investigate coherent structures of the most energetic modes in the wake of the cone flows. The velocity field data set is interpolated to a coarse mesh to reduce the computational cost. A 3D domain $(0\le x/D\le 20, -2.5\le y/D \le2.5, -2.5\le z/D \le2.5)$ with grid points of $160\times50\times50$ is used for the interpolation. We have performed a grid resolution study such that the small domain with the coarse mesh is sufficient to resolve the leading energetic POD modes for all the cases. We use 120 snapshots over convective time of $tU_\infty/L=4$ for all the cases in the POD analysis. This time interval covers approximately 5, 6, 7, and 9 periods of dominant frequency for the Reynolds numbers 500, 750, 1000, and 1500, respectively. Figure \ref{fig:energymode} shows the ratio ($\sum E_k$) of the sum of the energy of POD modes ($\sum_{j=1}^{k} \lambda_j$) to the total energy ($\sum_{j=1}^{n} \lambda_j$) of all the POD modes. For the flows at $Re$ = 500 that have clear large-scale hairpin-vortex coherent structures, the first three POD modes contain more than $80\%$ of the total energy for both flows over the smooth and annulated cones. As $Re$ increases, the downstream flow becomes more complex, requiring more modes to contain $80\%$ of the total energy. At $Re$ = 1500, nearly 25 and 40 modes are needed to represent $80\%$ of the total energy for the smooth and annulated cone flows, respectively.

\begin{figure} [hbpt]
    \centering
    \includegraphics[width=1\textwidth]{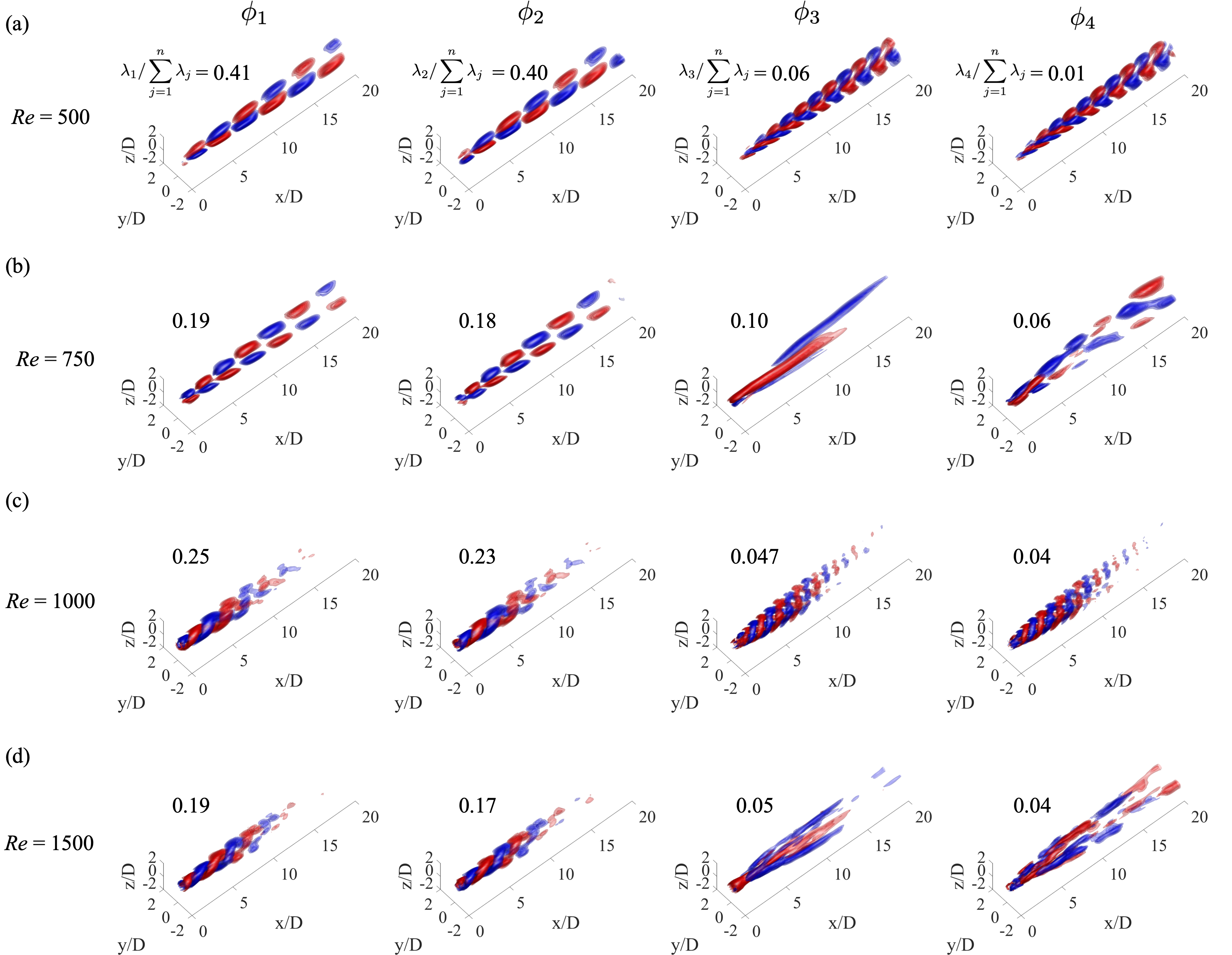}
    \caption{Iso-surface of streamwise velocity (red: $u/U_\infty\in[0.01,0.05]$, blue: $u/U_\infty\in[-0.05,-0.01]$) of the leading four POD modes ($\phi$) for the smooth cone flows at (a) $Re$ = 500, (b) 750, (c) 1000 and (d) 1500.}
    \label{fig:smoothmode}
\end{figure}

As the POD algorithm ranks the modes from the most energetic ones to the least energetic ones, the iso-surface of streamwise velocity $u$ of the leading four POD modes are plotted in figure \ref{fig:smoothmode}. At $Re=500$, the most energetic modes, the first and second POD modes ($\phi_1$ and $\phi_2$), have similar structures that the positive and negative disturbances alternate and align along the streamwise direction, representing the hairpin-vortex shedding feature as discussed in instantaneous flow fields (section \ref{sec:instant}); hence this type of mode is referred to as {\it hairpin-vortex mode}. The third and fourth POD modes (containing 6\% and 1\% mode KE, respectively) have similar structures as $\phi_1$ and $\phi_2$ but with smaller spatial scales, which agrees with the observation in the PSD analysis that strong harmonics are captured in the smooth cone flow at $Re=500$. However, as the flow's $Re$ is closer to the critical $Re$ that the shedding mechanism changes, the sub-dominant modes might not exhibit smaller-scale structures as harmonic modes (i.e., $\phi_3$ and $\phi_4$ of smooth cone flow at $Re=750$ in figure \ref{fig:smoothmode}(b)), but could represent a slow drift mode of the mean flow as the $Re$ is around the critical $Re$ of wake transition. Such modes are also observed in the annulated cone flows discussed later. In section \ref{sec:instant}, we have discussed that the flow shedding mechanism of the smooth cone changes when $Re\ge 1000$, and the spatial structures of the POD modes also depict this transition. As seen in figure \ref{fig:smoothmode}(c) and (d) at $Re=1000$ and 1500, respectively, the disturbances in the first and second POD modes twist and braid while convecting downstream, presenting a spiral structure around the cone axis, which is observed in the instantaneous spiral-vortex wake; hence this type of mode is referred to as {\it spiral-vortex mode}.

\begin{figure} [hbpt]
    \centering
    \includegraphics[width=1\textwidth]{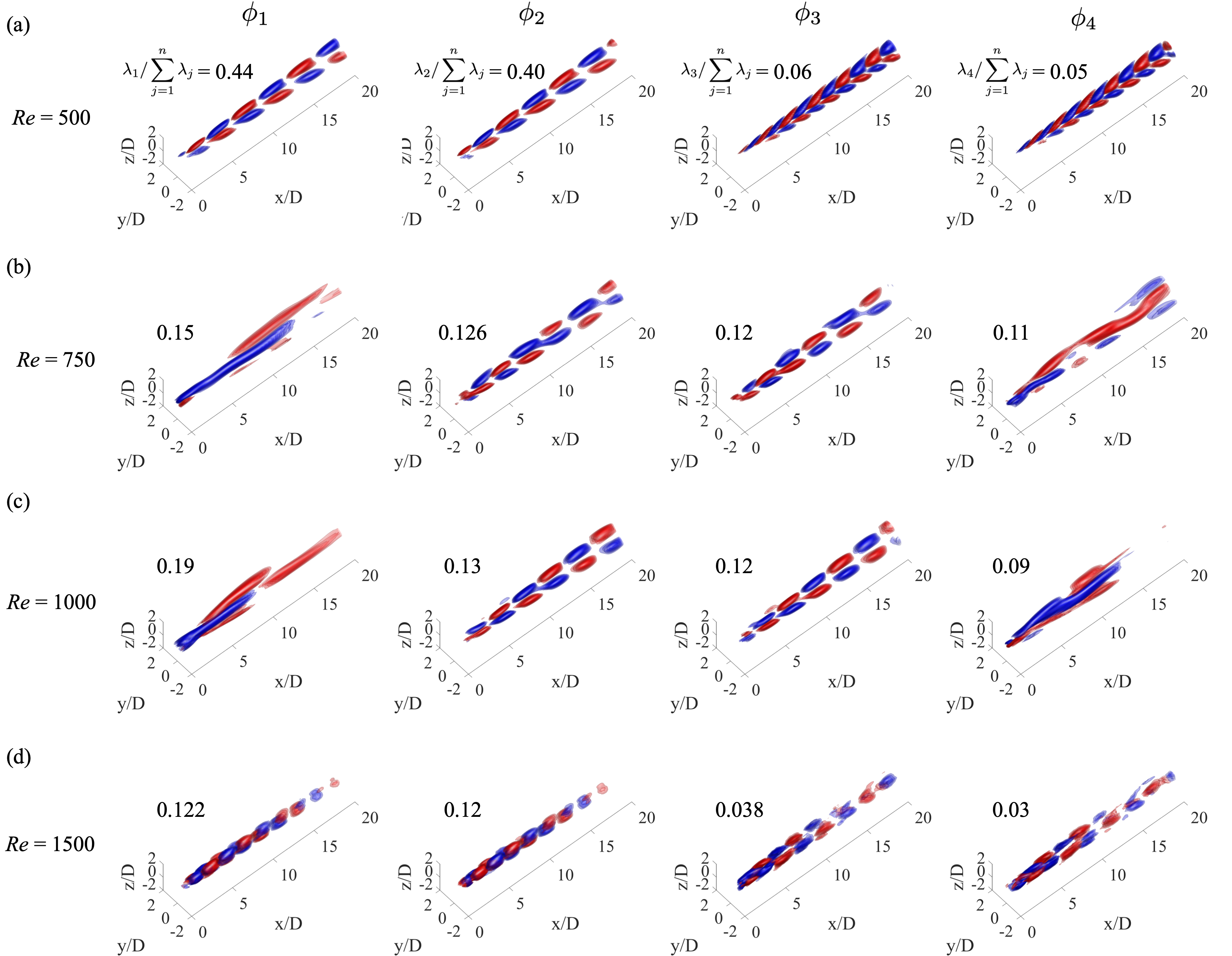}
    \caption{Iso-surface of streamwise velocity (red: $u/U_\infty\in[0.01,0.05]$, blue: $u/U_\infty\in[-0.05,-0.01]$) of the leading four POD modes ($\phi$) for the annulated cone flows at (a) $Re$ = 500, (b) 750, (c) 1000 and (d) 1500.}
    \label{fig:annulatedmode}
\end{figure}

The first four POD modes for the annulated cone flows are shown in figure \ref{fig:annulatedmode} at various Reynolds numbers. As the annulation essentially changes the boundary layer thickness and further yields a delay of transition of the shedding mechanism, we expect that the corresponding POD modes present similar features to the smooth cone flows. At $Re$ = 500, the wake remains hairpin-vortex wake so that the POD modes are similar to the ones captured in the smooth cone case (figure \ref{fig:smoothmode}(a)). Even at $Re=750$ and 1000, the leading POD modes still display features associated with hairpin-vortex wake. The elongated structures observed in the first modes for the annulated cone flows at $Re=750$ and 1000 could indicate a slow drift in the mean flow. A similar mode structure is observed in $\phi_3$ for the smooth cone case with lower energy in figure \ref{fig:smoothmode}. As such modes are observed around the critical $Re$, we speculate that such mode indicates a slow drift in the mean flow over the periods of the considered snapshots. Until the $Re$ increases to 1500, spiral-vortex modes appear as leading modes. However, the energy of these modes is comparatively lower than the smooth cone case at $Re=1500$. 

To investigate the coherent structures in the frequency domain, we further perform the SPOD analysis for the three selected cases, $Re=500$ and 1000 for the smooth cone flows and $Re = 1000$ for the annulated cone flow, due to their different vortex shedding mechanisms present in the wake. In the above analysis, the hairpin-vortex wake and spiral-vortex wake are observed for the smooth cone flows at $Re = 500$ and $1000$, respectively (figure \ref{fig:isoQ_smooth}), whereas the annulated cone flow at $Re=1000$ displays a mixture of the hairpin-vortex and spiral-vortex shedding mechanisms. The SPOD analysis considers 1024 snapshots with an equal temporal interval for all three cases. For the smooth cone cases, the number of snapshots considered in each block is 128 and 200, with a 75\% of overlap for $Re=500$ and $1000$, respectively. The total numbers of blocks are 29 and 19 in the corresponding cases. For the annulated cone flow at $Re=1000$, 128 snapshots are considered for each block with a 75\% of overlap, and the total number of blocks is 29. We have verified the number of snapshots per block and the total number of blocks to ensure the convergence of the solution \cite{schmidt2020}.

\begin{figure} [hbpt]
    \centering
    \includegraphics[width=0.9\textwidth]{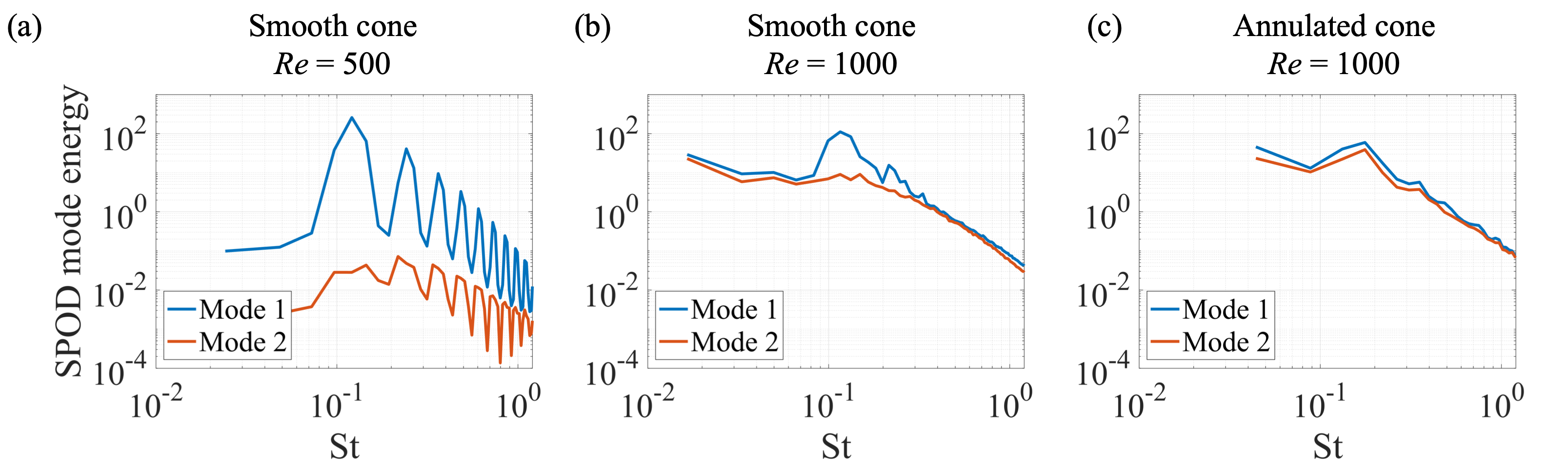}
    \caption{The SPOD mode energy spectrum of the first and second modes. (a) Smooth cone flow at $Re = 500$, (b) smooth cone flow at $Re = 1000$, and (c) annulated cone flow at $Re = 1000$.}
    \label{fig:SPOD_energy}
\end{figure}

The SPOD eigenvalue spectrum of the streamwise velocity for the three cases is shown in figure \ref{fig:SPOD_energy}. At each frequency, if the leading SPOD mode contains energy much larger than the second mode, the flow is considered to have rank-1 behavior. Here, only the first two modes are shown for the eigenvalue spectrum to identify the rank-1 behavior of the flows.  For the smooth cone flow at $Re = 500$ (figure \ref{fig:SPOD_energy}(a)), the energy difference between the first and second modes is significant for all the frequencies. It indicates that the physical mechanism associated with the first mode is prevalent, and the flow exhibits a rank-1 behavior. We also note that the POD analysis reveals a low-rank behavior for this case, in which the first two POD modes contain nearly 80\% of the total energy, although it is not rank-1 due to the comparable energies contained in modes 1 and 2. At $St=0.12$, a peak in the energy spectrum is observed, and the other peaks are harmonic frequencies to the dominant one. For the smooth cone flow at $Re = 1000$, the rank-1 behavior is observed over $0.082 \lesssim St \lesssim 0.198 $. The peak in the energy spectrum is observed at $St=0.11$, which is within the range of the rank-1 behavior. However, the annulated cone at $Re=1000$ exhibits a very close energy spectrum of the first and second modes (figure \ref{fig:SPOD_energy}(c)), and the assumption of the rank-1 behavior is invalid for this case where a mixture of shedding mechanisms are captured. The peaks in the energy spectrum for the first and second modes are at $St = 0.17$, which is similar to the dominant frequency observed from the PSD analysis.

\begin{figure} [hbpt]
    \centering
    \includegraphics[width=1\textwidth]{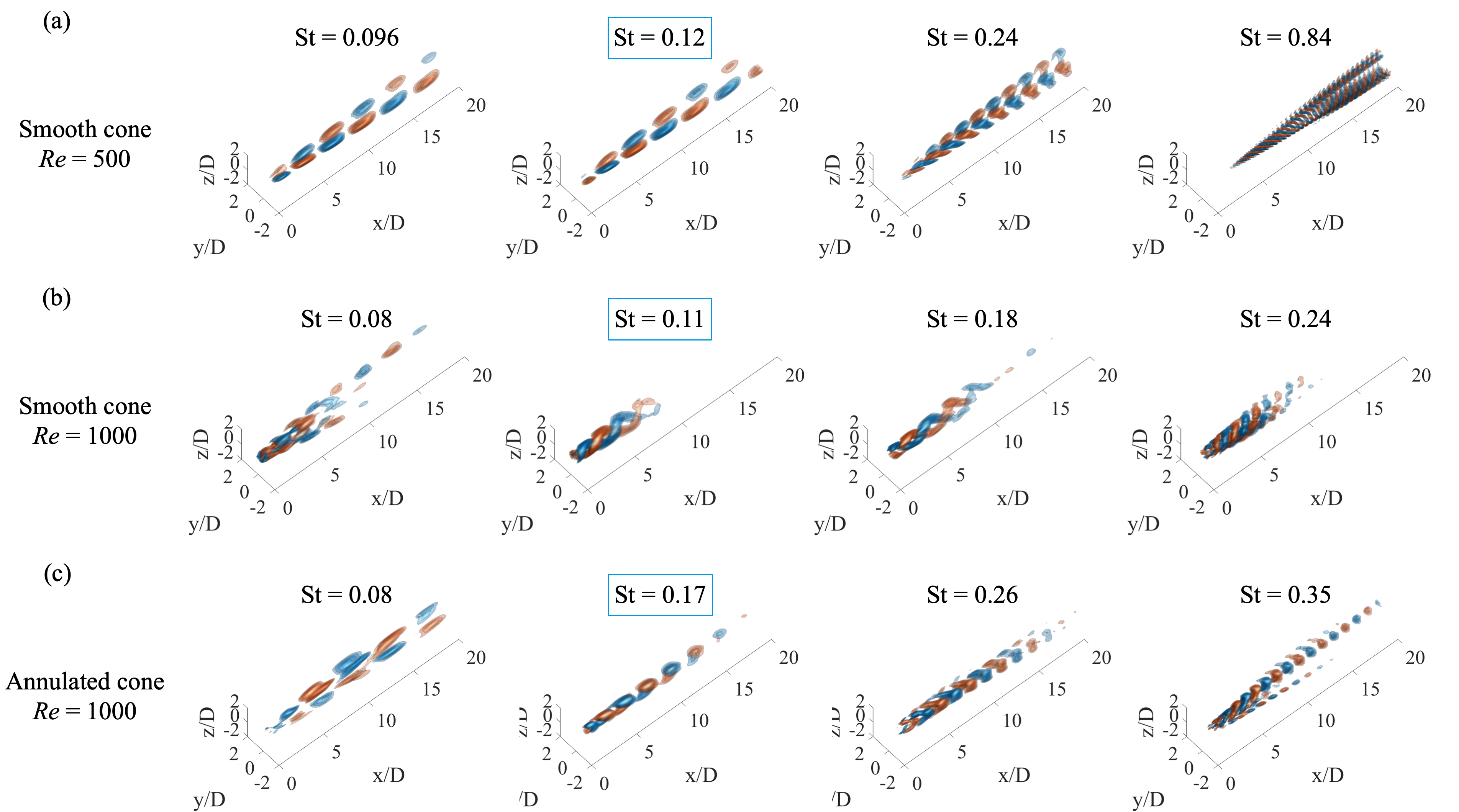}
    \caption{Iso-surface of streamwise velocity (red: $u/U_\infty\in[0.01,0.05]$, blue: $u/U_\infty\in[-0.05,-0.01]$) of the leading first SPOD modes at different frequencies. (a) Smooth cone, $Re = 500$, (b) Smooth cone, $Re = 1000$, and (c) Annulated cone, $Re = 1000$.}
    \label{fig:SPODmode}
\end{figure}

The leading SPOD modes of these three cases at different frequencies are shown in figure \ref{fig:SPODmode}. The frequencies highlighted by blue boxes are the peak frequencies captured in the energy spectrum of the SPOD analysis. For the smooth cone flow at $Re=500$, the SPOD mode with $St=0.12$ displays a spatial structure of alternately positive and negative disturbances similar to the leading POD mode (figure \ref{fig:smoothmode}(a)), representing the hairpin-vortex wake. The modes associated with higher frequencies of $St=0.24$ and 0.84 have smaller-scale structures and are closely packed downstream. The POD and SPOD analyses indeed capture resemblant mode shapes, while the SPOD analysis provides additional information about the frequency associated with those prominent modes. For the smooth cone flow at $Re=1000$, the leading mode at frequency $St=0.11$ displays a spiral shape, and the modes with higher frequencies ($St = 0.18$ and 0.24) have a similar shape but with smaller-scale structures. Moreover, the leading SPOD mode at all frequencies exhibits the twisting feature in space which emphasizes the rotating behavior of the wake flow for the smooth cone at $Re=1000$. For the annulated cone at $Re=1000$, the leading mode at $St=0.17$ contains the twisting-shape mode; however, based on a qualitative comparison of the mode shape, the twisted effect is not as strong as observed in the smooth cone flow at the same Reynolds number. At the lower frequency $St=0.08$ and higher frequency $St=0.26$, the SPOD modes display the hairpin-vortex mode structures similar to the modes observed at the low Reynolds number of $Re=500$. Hence, for the flow with a mixed shedding mechanism, the SPOD analysis can reveal the two different shedding mechanisms in the wake, whereas the POD analysis can only capture the hairpin-vortex modes as leading modes.

\section{Summary}
\label{summary}

This study examines annulation effects on a flow over an orthoconic structure by performing direct numerical simulations at various Reynolds numbers. We observe two different shedding mechanisms for the cone flows. At a low Reynolds number ($Re\le 750$), vortical streaks develop from the cone end, and hairpin-shaped vortical structures grow as the flow convects downstream (referred to as hairpin-vortex wake). At a high Reynolds number ($Re\ge1000$), the vortical structures shed in a spiral pattern from the cone end with small-scale turbulent structures around the large-scale coherent structures (referred to as spiral-vortex wake). The flow's shedding mechanism can be changed by introducing annulations to the cone surface, in which the appearance of the spiral-vortex wake can be delayed by adding surface annulation. Moreover, the spectral analysis shows that the dominant shedding frequency increases substantially because of the change in the shedding mechanism in the wake. Compared to the smooth cone, the dominant frequency decreases marginally in the annulated cone cases. 

The time-averaged streamlines indicate that the recirculation zone after the cone end is asymmetric to the cone axis if the shedding mechanism is the hairpin-vortex wake. When the Reynolds number increases, the shedding mechanism becomes a spiral-vortex wake, and the recirculation zone will become axisymmetric to the cone axis. The drag force increases in the annulated case due to a drastic rise in the pressure drag. Although the viscous drag of the annulated cone flow reduces due to the flow trapped in the groove of the annulation creating a slip condition for external flow, the reduction in the viscous force is minimal compared to the increase in pressure drag. The instantaneous pressure force at the cone end shows the pressure force direction is almost invariant if the hairpin-vortex wake dominates the wake flow. In contrast, the pressure force at the cone end will rotate based on the azimuthal-shedding direction of the spiral-vortex wake. Furthermore, the annulation over the cone surface reduces the pressure fluctuation in the wake compared to the smooth cone flows. 

In the modal decomposition analysis (POD and SPOD analyses) to uncover the dominant coherent structures, we find that the leading POD modes can either capture hairpin-vortex modes or spiral-vortex modes in each flow. However, the SPOD analysis can successfully capture both modes and isolate them depending on their frequencies in the flow that contains both shedding mechanisms (annulated cone flow at $Re=1000$). Hence, the SPOD analysis is more suitable for analyzing the flow's wake transition where the shedding mechanism is complex. In the present study, we have examined the effects of annulation of the cone flow and found that adding surface annulation can delay the transition of the wake from hairpin-vortex wake to spiral-vortex wake. This mechanism can shed light on morphological surface design in the engineering applications of small underwater vehicles or other orthoconic structure devices at low and moderate Reynolds numbers.    

\section*{Acknowledgments}
This work is supported by the Collaboration for Unprecedented Success and Excellence and Excellence Grant at Syracuse University and the Syracuse Office of Undergraduate Research and Creative Engagement.  We acknowledge the Research Computing Center at Syracuse University providing computational resources. We also thank the New York State Museum for a loan of specimens from the Invertebrate Paleontology collection.

\bibliography{references}
\bibliographystyle{acm}

\end{document}